*Children's expressed emotions during playful learning games*

Thomas Vase Schultz Volden and Paolo Burelli
Digital Design, Center for Digital Play, BrAIn Lab, IT University of Copenhagen, Denmark
thvo@itu.dk
pebu@itu.dk

**Abstract:** Studies on software tutoring systems for complex learning have shown that confusion has a beneficial relationship with the learning experience and student engagement (Arguel et al., 2017). Causing confusion can prevent boredom while signs of confusion can serve as a signal of genuine learning and as a predecessor for frustration. There is little to no research on the role of confusion in early childhood education and playful learning, as these studies primarily focus on high school and university students during complex learning tasks.

Despite that, the field acknowledges that confusion may be caused by inconsistency between information and a student's internal model referred to as cognitive disequilibrium known from the theory of cognitive development, which was originally theorized based on observational studies on young children (D'Mello and Graesser, 2012). Therefore, there is reason to expect that the virtues of confusion also apply to young children engaging in learning activities, such as playful learning.

To investigate the role of confusion in playful learning, we conducted an observational study, in which the behavior and expressed emotions of young children were collected by familiar pedagogues, using a web app, while they engaged with playful learning games designed for kindergartens. The expressed emotions were analyzed using a likelihood metric to determine the likely transitions between emotions (D'Mello and Graesser, 2012).

The preliminary results showed that during short play sessions, children express confusion, frustration, and boredom. Furthermore, the observed emotional transitions were matched with previously established models of affect dynamics during complex learning. We argue that games with a learning objective can benefit by purposely confusing the player and how the player's confusion may be managed to improve the learning experience.

## 1. Introduction

Assessing the educational benefits of playful learning can pose challenges. Play is inherently voluntary, meaning participants can shape the context around their engagement often in a frivolous manner (Sutton-Smith, 2001). Consequently, planning and evaluating a curriculum around play becomes complex, as it arguably gauges participants' willingness to adhere to it rather than the effectiveness of the curriculum itself. Gathering data from early educational games utilizing playful learning strategies and comparing such data with studies on learning, in general, may produce some evidence for the effect of playful learning. Not to mention provide some guidance for future endeavors in playful learning games.

Studies on Academic emotions have shown that students engaging in learning activities, experience affective states categorized as epistemic emotions (Pekrun and Linnenbrink-Garcia, 2012). Affective state is a term used to describe emotions and moods more specifically as a felt affect with a positive or negative valence (Feldman Barrett and Russell, 1998). Positive affective states such as enjoyment and negative affective states such as frustration and boredom are shown to be related to study activities (Pekrun and Linnenbrink-Garcia, 2012). This study set out to investigate if young children express the same emotions and epistemic affective states during playful learning games for their age group as university students have been observed expressing during complex learning. In particular, studies that model transitions between emotions, called model on affect dynamics, have proved effective as a guide for automated tutoring software as they only have to detect and address certain emotions to optimize the learning experience (D'Mello and Graesser, 2012). To investigate this, the following research question was formulated.

*Do young children experience academic emotions such as confusion, frustration, and boredom while engaging in playful learning games and do these emotions adhere to the model on affect dynamics?*

Emotions are an integral part of humans and an important factor for certain activities such as learning, where confusion, frustration, and boredom have been shown to have positive and negative effects on deep learning (D'Mello and Graesser, 2014). Furthermore, studies of the model on affect dynamics have shown that students tend to transition between engagement ↔ confusion ↔ frustration ↔ boredom, which means that students do not go directly from engaged to frustrated without passing through a stage of confusion (D'Mello and Graesser, 2012). Therefore, detecting and acting on confusion may alleviate frustration and prevent boredom. Other studies suggest that engagement can transition directly to boredom and can be prevented by intentionally causing confusion (Arguel et al., 2019b).

The model on affect dynamics has been supported by multiple accounts (Arguel et al., 2019a; Mulqueeny et al., 2015; Munzar et al., 2021). The original studies used self-reporting during a short (30 min) individual interaction with a program called AutoTutor which is developed to teach complex subjects via dialogue (D'Mello and Graesser, 2012). Participants had to select an emotion or epistemic affective state from a list: Engagement/flow, confusion, frustration, boredom, anxiety, curiosity, delight, and surprise. More recent studies are based on one or more observers' frequent observations of the individual participant's activities and expressed emotions or epistemic affective states (Baker et al., 2020). Activities can be either on task, on task while having a conversation, off task, or using the system in a way that is not intended or unknown. The expressed emotions or epistemic affective states are reduced to boredom, engagement, confusion, frustration, or unknown. These studies have been developed into a standardized protocol called BROMP (Ocumpaugh et al., 2015) associated with a certification program and a phone app to record observations. The protocol can be utilized for single participants or multiple participants interacting with a system simultaneously, by observing each participant in the group in iterations. Although teachers are expected to possess the ability to discern confusion among students and respond appropriately (Arguel et al., 2017), BROMP emphasizes the necessity of training to consistently generate accurate observations (Ocumpaugh et al., 2015). Unfortunately, the studies of the model on affect dynamics focus mainly on students learning complex subjects, such as mathematics, in higher education (Arguel et al., 2017). The role of confusion in early childhood education and what can be considered equivalent complex learning for young children remain almost scientifically unexplored.

To get an idea of how confusion and complex learning affect early childhood education, we can examine the term complex learning in general. Complex learning is often exemplified as something that causes cognitive disequilibrium (D'Mello and Graesser, 2012). Cognitive disequilibrium can be caused by information that does not fit with an internal conceptualization of the world or an impasse that cannot be resolved with the current set of skills (Ormrod, 2019). Studies on confusion have shown a correlation between cognitive disequilibrium and confusion (D'Mello and Graesser, 2014, 2012). Studies of young children have documented that children experience cognitive disequilibrium around the age of 6 years (Ormrod, 2019). These studies observed that children were intrinsically motivated to reestablish cognitive equilibrium, which led them to develop more complex levels of thought and knowledge, which suggests that young children also benefit from confusion management strategies (Arguel et al., 2019b).

This tells us that confusion may be as important for early childhood education as it is for complex learning, but the question remains, how can we utilize this knowledge? We may have to turn the problem on its head and ask if this knowledge can be used to optimize already established educational tools, such as playful learning. To do this, four hypotheses were formulated, which will show if young children engaging in playful learning adhere to the model on affect dynamics.

- H1. Young children engaging in playful learning transition from concentration to confusion.
- H2. Young children engaging in playful learning transition from confusion to concentration and not directly from frustration to concentration.
- H3. Young children engaging in playful learning transition from confusion to frustration and not directly from concentration to frustration.
- H4. Young children engaging in playful learning transition from frustration to boredom and not directly from confusion to boredom.

## 2. Review

Since gaining widespread popularity, digital games have captivated researchers for their teaching potential (Greenfield, 2014). Researchers have studied how young players were able to learn complex behaviors using deduction to master a game (Gee, 2007; Greenfield, 2014). The idea of presenting curriculum learning through games like Edutainment seemed apparent (Klawe and Phillips, 1995) but, in reality, turned out to be more complicated than expected (Arnseth, 2006; Egenfeldt-Nielsen, 2006).

More recent research on game-based learning has split up into different approaches, some of the promising being Serious Games (Zhonggen, 2019), where a game is based around an educational context, and Gamification (Sailer and Homner, 2020) where game elements are introduced to non-game related tasks such as learning. While Gamification primarily focuses on the motivational aspect of games (Mekler et al., 2017), studies have shown that ease of play and clear instructions are important factors for player engagement in serious games (Iten and Petko, 2016; Zhonggen, 2019). This aligns with the Flow theory which stresses the importance of balancing difficulty with player competence along with clear communication of goals and feedback (Csikszentmihalyi, 2014). Flow theory focuses mainly on experience and how to optimize it, identifying flow as a favorable experience while more negative emotions such as frustration and boredom should be avoided by adapting difficulty to the competence of the player. The difficulty adaption aspect of Flow theory has been studied in game settings, by adjusting difficulty based on performance (González-Duque et al., 2020; Kristensen et al., 2022) or based on player experience (Yannakakis and Togelius, 2011). However, a recent review questions the efficacy of Dynamic Difficulty Adjustment (DDA) techniques which aim to merely archive Flow (Guo et al., 2024) and suggest assessing both objective (performance) difficulty and subjective (perceived) difficulty to adjust game tasks in alignment with design goals. For example, an educational game can impact the effect of learning with objective performance while subjective difficulty is determined by player experience.

Studies on confusion during complex learning may prove useful in assessing player experience. Confusion has been identified as a signal for genuine learning (D'Mello et al., 2014) while the experience of confusion can be negative or positive depending on how it is managed (Arguel et al., 2019a). In one study, students were assigned to solve a particularly unintuitive and frustrating puzzle called Lights Out, the study showed that providing the students with a strategy positively changed the experience of confusion (Arguel et al., 2019a). Hence the experience of confusion may be influenced by a student' available strategies resulting in productive confusion or hopeless confusion and eventually enjoyment or frustration (D'Mello and Graesser, 2012). Furthermore, confusion has been shown to precede frustration (D'Mello and Graesser, 2012) and as a means to prevent boredom (Arguel et al., 2019b). Arguably player confusion may be a suitable guide for subjective difficulty in learning games and may be managed by providing the player with helpful strategies. This may be particularly important for early childhood educational games, as young children may not have accumulated many problem-solving or emotional management strategies.

## 3. Study

The objective of the study was to observe young children expressed affective states during relatively complex playful learning games. Pedagogues familiar with the children were deemed suited to recognize expressed emotions based on feedback from professional pedagogue educators and practitioners. Analysis of the observations determines the likelihood of one emotion transitioning to another, which is used to prove or disprove the hypotheses and provide an answer to the research question.

### 3.1 Procedure

Participants played three selected games in random order. For each game, the participant was briefly instructed on how to play, after which the participant had seven minutes to find as many unique solutions as possible. While the participant was engaged in the activity, a pedagogue who was acquainted with them observed and documented their behavior and affective state. To collect observations, a mobile-friendly web app was developed by following the specifications of the mobile application used in BROMP (Ocumpaugh et al., 2015). Before the first play session, the pedagogue was instructed on how to use the web app, randomize the game order, and not intervene with the activity unless the participant gave up entirely or requested help.

## 3.2 Material and equipment

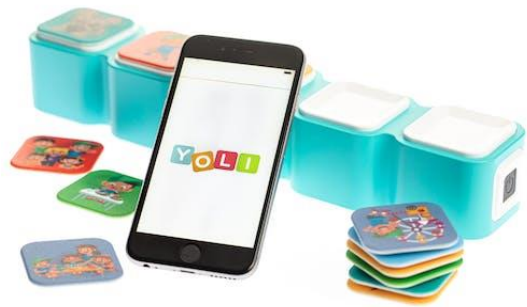

**Figure 1:** The digital YOLI board, some YOLI game tiles, and a phone with the YOLI app running (picture from playyoli.com).

YOLI has developed a digital platform for tile-based games, see **Figure 1**. YOLI Games comes in sets of thirty tiles featuring a picture on each and an RFID chip with information about the tile. The YOLI board has room for five tiles, each fitted with an RFID reader and an electromagnet which allows it to shake or even throw off tiles that are placed on it. A smartphone can connect to the board and extend the possibilities using the YOLI app, which will also fetch anonymous activity logs from the game and report them to YOLI.

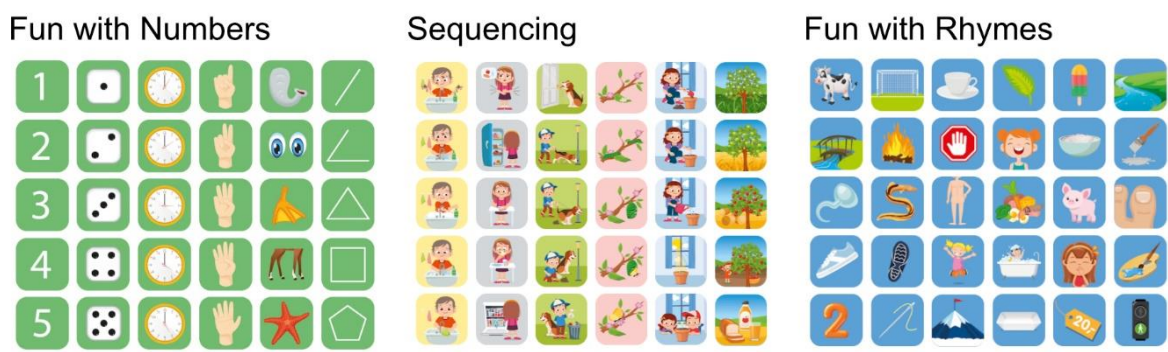

**Figure 2:** Overview of all tiles for the three selected games, where tiles are placed in the "correct" order top-down and horizontally for the Fun with Numbers game.

The goal of a YOLI game is commonly to place five tiles on the board following the rule of the game, which is normally to match the different tiles on one or multiple parameters. For the study, three games were selected in collaboration with YOLI and feedback from pedagogues. We aimed to select games with diverse learning areas to raise the likelihood of having at least one game which all participants would find harder than the others.

The games that were selected, shown in Figure 2, were: Fun with Numbers, Sequencing, and Fun with Rhymes.

**Fun with Numbers**

The goal is to place five tiles on the board either by their numeric value or in numerical order for a specific symbol. An example of the same numeric value for multiple symbols is the top horizontal line for Fun with Numbers in **Figure 2** which shows: 1, one rolled on a six-sided die, a watch showing one o'clock, a single finger, an elephant trunk, and a single line. Combining five of these different symbols with the same numeric value in any order will complete the goal.

This game was chosen because of its emphasis on numbers and the ability to identify the numeric value of symbols.

**Sequencing**

The goal is to arrange five tiles into the correct sequences. The sequences are: Washing dirt off your hands using soap, reacting to hunger by making a sandwich and cleaning up after it, taking the dog for a walk and cleaning up after it, the life cycle of a butterfly, planting and growing a seed, the seasonal cycle of apples and wheat ending with consumer products.

This game was chosen because of its diversity in sequence, which requires some knowledge of multiple subjects plus the ability to identify the small clues in the images.

**Fun with Rhymes**

The goal is to place five tiles on the board that rhyme, this is particularly tricky as it may not be apparent to the child which word is associated with the image on the tiles. Under normal circumstances a pedagogue would help identify the word, but for the study pedagogues were asked not to help unless the child explicitly asked for help.

### 3.3 Participants

Recruitment attempts on social media groups specifically targeting pedagogues did not provide any interest. Directly contacting twenty kindergartens and after-school daycare centers provided 3 pedagogues/institutions willing to participate. Participants were selected by the pedagogies of the willing institutions.

The participants were 10 children, seven from kindergarten and three from after-school daycare. There were four five-year-olds and six six-year-olds. Seven of the participants were male and three were female. Informed consent forms were gathered from parents by the pedagogues and no reward was provided for participating.

### 3.4 Data analysis

The study produced 1113 observations with an average of 37 (SD = 8) observations per game, meaning that the pedagogues recorded observations every 12 s (SD = 9) per average. There were 810 observations of concentration, 175 confusion, 45 frustration, 17 boredom, and 66 unknown.

The analysis for this study followed the procedure from the original study on transitions between epistemic affective states (D'Mello and Graesser, 2012). First, redundant observations of the same affective state (X -> X -> Y) were removed so that transitional observations were left (X -> Y), this reduced the observations to 220 with an average of 7 (SD = 5) transitions per game, with a transition every 51 s (SD = 55) on average.

$$L(M_t \rightarrow M_{t+1}) = \frac{P(M_{t+1}|M_t) - P(M_{t+1})}{1 - P(M_{t+1})} \qquad (1)$$

A likelihood metric was produced to compute the likelihood of a transition between two affective states (see equation (1)) where $M_t$ is the current affective state and $M_{t+1}$ is the next affective state (D'Mello and Graesser, 2012). The likelihood is calculated by taking the conditional probability $P(M_{t+1}|M_t)$ of the next affective state being $M_{t+1}$ if the current affective state is $M_t$, this is then subtracted by the probability $P(M_{t+1})$ of the next affective state being $M_{t+1}$ by chance. By dividing these probabilities by the probability of $M_{t+1}$ not being the next affective state, we get the association between affective states. If the current and next affective states are positively associated, the result will be positive, indicating that $M_t$ is likely to transition into $M_{t+1}$, a negative value indicate the opposite, that $M_t$ is not likely to transition to $M_{t+1}$. A result of zero indicates no association between the two states. The probabilities were calculated by counting the number of next affective states observed divided by the total number of observations. For the conditional probability the number of next affective state was divided by the number of observations for each of the different affective states.

### 3.5 Ethics statement

The study has been discussed with the local ethics council of the university as well as pedagogical researchers. There was an emphasis on the experience of the child during and after play, which should result in the child leaving the session with an uplifted feeling, therefore the study was presented as an opportunity for the child to help improve the games, followed by some uplifting comment and asking if the child had fun and acknowledging their helpful feedback. Informed consent forms were formulated for the parents of the participants and cleared by the legal department.

### 3.6 Results

The likelihood was calculated for every conceivable shift in affective states, using equation (1), resulting in 20 possible transitions within a 5x4 matrix. Out of the 20 possible transitions, seven were chosen as the ones that could prove or disprove the hypotheses. The transition from concentrating to bored was not included since the original study on affect dynamics as well as later studies have shown that participants can go directly from concentrating to bored (supposedly because of a lack of confusion) (Arguel et al., 2019b; D'Mello and Graesser,

2012). Four of the transitions are directly related to hypotheses, those are Concentrating → Confused related to H1, Confused → Concentrating related to H2, Confused → Frustrated related to H3, and Frustrated → Bored related to H4. Three transitions were selected as their positive association would contradict the hypotheses, those were Frustrated → Concentrating to contradict H2, Concentrating → Frustrated to contradict H3, and Confused → Bored to contradict H4. A one-sample *t*-test was applied to test whether there was a significant likelihood of association between a current- and a next affective state. While all participants were observed as concentrating, two participants did not experience confusion, two did not experience frustration, and seven did not express boredom. Five participants were noted as expressing emotions that were not among the pre-defined ones at one point during the play session. One participant was only observed as concentrating during the entire session.

Descriptive statistics on the likelihood of the transitions and the results of the *t*-test are presented in **Table 1**, where the average is calculated between game trials (3 per participant) and the N-column is the number of trials where the given transition was observed.

**Table 1:** Transition likelihood for select transitions.

| Transition | Descriptives | | | One-sample *t*-test | | |
|---|---|---|---|---|---|---|
| | N | Mean | SD | t | df | p |
| Excitatory | | | | | | |
| Concentrating → Confused* | 17 | .562 | .484 | 5.07 | 18 | .000 |
| Confused → Concentrating | 17 | .312 | .835 | 1.87 | 24 | .074 |
| Confused → Frustrated | 4 | -.145 | .526 | -.995 | 12 | .339 |
| Frustrated → Bored | 1 | -.009 | .319 | -.061 | 4 | .954 |
| Inhibitory/baseline | | | | | | |
| Frustrated → Concentrating* | 8 | -.328 | .772 | -2.13 | 24 | .044 |
| Confused → Bored | 1 | -.138 | .216 | -1.43 | 4 | .227 |
| Concentrating → Frustrated* | 7 | .283 | .389 | 2.62 | 12 | .022 |
| No prediction | | | | | | |
| Concentrating → Bored | 4 | .113 | .140 | 1.80 | 4 | .146 |
| Frustrated → Confused | 5 | -.385 | .867 | -1.93 | 18 | .069 |
| Bored → Confused* | 1 | -.776 | .443 | -7.65 | 18 | .000 |
| Bored → Concentrating* | 4 | -.566 | .645 | -4.39 | 24 | .000 |
| Bored → Frustrated* | 1 | -.316 | .356 | -3.20 | 12 | .008 |

Notes. *p < .05.

The results show that hypothesis 1, which predicts the transition from concentrating to confused was supported by a significant positive association. There is some support for hypothesis 2, which predicts a transition from confused to concentrating and a significant negative association between frustrated and concentrating. Hypothesis 3 is rejected by a negative association between confused and frustrated and a significant positive association between concentrating and frustrated. The weak negative association for the transition from frustrated to bored contests hypothesis 4, whereas the negative association for the transition from confused to bored supports the constraint, none of them significant. This indicates insufficient data to prove or disprove hypothesis 4.

Additionally, there appears to be a significant negative association between boredom and any of the other three pre-defined affective states, however, these were only observed in very few trials.

*3.6.1 Emotion Assessment*

While all three pedagogues agreed that a pedagogue who is familiar with the participant is best suited to accurately observe emotions in the participant, two of the pedagogues did express difficulty with confusion and frustration. Arguably, confusion is not an emotion that pedagogues are as acquainted with as teachers, and it can be hard to distinguish between confusion and frustration. When asked the pedagogues explained that they mainly look for confusion in the eyes of the participants. Studies have associated facial expressions like lowered brows and tightened eyelids as indicators of confusion (D'Mello et al., 2014). This supports the claim that observers should receive some training in recognizing confusion and frustration (Ocumpaugh et al., 2015).

## 4. Discussion

The study was originally designed as two parts, one was to video record children playing the game and then at a later point have multiple pedagogues individually annotate the same videos with labels of emotions. The alignment between pedagogue annotations, plus an additional automated facial expression detection algorithm would provide data for the affective state transitions. This would not only provide evidence for the study but also allow for a critical analysis of automated facial expression detection of young children. The study was redesigned into the final observational study, as we were unable to recruit pedagogues to annotate videos. The pedagogues did not fully condone video recording children in institutions and argued on multiple occasions independent of each other, that emotional recognition could only reliably be done by a pedagogue that is familiar with the child. The pedagogues who conducted the observations agreed with this argument, even though they found it difficult to recognize the confusion and frustration, also in situations where the child's behavior could indicate confusion and when the child actively asked the pedagogue to help. Upon reflection, pedagogues agreed that the child must have been confused, however, they did not observe that the child explicitly expressed confusion or frustration in the situation. Since the study relied on pedagogues' ability to recognize confusion, frustration, and boredom, it is relevant to discuss: 1) based on their experience with the children they care for are pedagogues suited to detect complicated emotions such as confusion and frustration? 2) should complicated emotions such as confusion and frustration and how to manage them, be a focus of pedagogues? Arguably, more data, than this study has produced, is needed to enrich such a discussion.

Another aspect that the researchers have taken note of, is the reluctance of pedagogues and pedagogical institutions to participate in evidence-based research. While the reason for this reluctance is outside the scope of the study, it did have a direct consequence on the study's validity due to the scarcity of data. Furthermore, researchers encountered multiple accounts of pedagogues who independently of each other questioned the practice of this study to produce evidence. Arguably the evidence is considered superficial due to the unnatural environment of the study, where children are placed in unfamiliar circumstances and asked to perform a task.

## 5. Conclusion and future research

This study set out to find evidence that young children experience the same academic emotions during playful learning games, as students have been observed during complex learning tasks. To do this, we observed children play a selection of playful learning games, which can already be found in some kindergartens today. Professional pedagogues familiar with the child observed the expressed affective states during play. The results showed that young children do experience confusion, frustration, and boredom during playful learning. The results were further analyzed to check if the transition between emotions follows the model on affect dynamics, which states that students tend to transition from concentration, through confusion to frustration and eventually boredom. To answer this, four hypotheses were formulated and tested. The result showed that young children do tend to go from concentrating to confusion when engaging in playful learning. However, the evidence also showed that there was a tendency to transition from concentration to frustration directly. Therefore, we are unable to prove that young children adhere to the model on affect dynamics, which we speculate is due to a lack of training of the observers. Furthermore, the results were insufficient to prove or disprove whether young children are inclined to transition from frustration to boredom and not directly from confusion to boredom.

While we can say that children experience these emotions, more research is needed to reliably say if young children adhere to the model of affect dynamics during playful learning games.